\documentclass[5p,sort&compress]{elsarticle}
\usepackage{graphicx}
\usepackage{amsmath}
\usepackage{hyperref}

\begin{document}

\title{Strain-induced pseudomagnetic and scalar fields in symmetry-enforced Dirac nodes}
\author[VNIIA]{A. D. Zabolotskiy}
\ead{zabolotskiy@vniia.ru}
\author[VNIIA,ISAN,HSE]{Yu.\ E. Lozovik}
\ead{lozovik@isan.troitsk.ru}
\address[VNIIA]{Dukhov Research Institute of Automatics (VNIIA), Sushchevskaya~22, 127055 Moscow, Russia}
\address[ISAN]{Institute of Spectroscopy of the Russian Academy of Sciences (ISAN), Fizicheskaya~5, 108840 Troitsk, Moscow, Russia}
\address[HSE]{Moscow Institute of Electronics and Mathematics, National Research University Higher School of Economics (MIEM~HSE), Tallinskaya~34, 123458 Moscow, Russia}

\begin{abstract}
It is known that Dirac nodes can be present at high-symmetry points of Brillouin zone only for certain space groups. For these cases, the effect of strain is treated by symmetry considerations. The dependence of strain-induced potentials on the strain tensor is found. In all but two cases, the pseudomagnetic field potential is present. It can be used to control valley currents.
\end{abstract}


\maketitle

\section{Introduction}
Dirac and Weyl semimetals, gapless materials with linear dispersion of low-energy electron excitations effectively described by Dirac-like or Weyl-like equation, are being extensively studied both theoretically and experimentally \cite{DiracReview14,DSMAnnuRev,DSMJPCM,NewRev}.

Elastics strain has a remarkable effect on Dirac and Weyl fermions in solids, initially found in 2D Dirac semi\-metal graphene: it induces scalar potential and vector (to be more precise, pseudovector) potential having exactly the form of potential induced by external magnetic field \cite{pmfreview,dsmpmf}. It is called pseudomagnetic field (PMF); the only difference between PMF and real magnetic field is that PMF does not break time reversal symmetry. The pseudomagnetic field can be very strong \cite{pmfexp}, but it is an effective field which is felt only by Dirac fermions and cannot be detected from the outside directly. Similar effect has been noted in materials other than graphene, too \cite{StrainRev,pmfreview}, see also earlier work~\cite{IK85}.

Thus strain can lead to the physics of magnetism even in non-magnetic materials such as graphene, where strain in combination with other stimuli could lead to diamagnetism, paramagnetism, or ferromagnetism~\cite{Ferro,YaB}.

Some conclusions about the existence of Dirac or Weyl nodes at high-symmetry points of Brillouin zone (BZ) can be made solely from the symmetry considerations \cite{NewRev,Manies12,symmDirac2D}. Ref.~\cite{Manies12} provides the corresponding data for ``orbital Weyl points'', i.~e., Weyl points in the band structure of spinless electrons, so these points are actually Dirac points (if we take into account spin degeneracy but neglect spin-orbit interaction). It is shown that some space groups enforce the presence of Dirac nodes at some high-symmetry points of BZ, others allow but not enforce it, and the rest forbid Dirac nodes at high-symmetry points. Even in the first case the resulting Dirac material is not necessarily a Dirac semimetal because Dirac excitations may coexist with some other electron excitations away from high-symmetry points.

It is naturally to expect the presence of strain-induced potentials in symmetry-enforced Dirac nodes. Conclusions about general analytical form of the strain-induced contribution to Dirac Hamiltonian can also be made on the symmetry grounds only, which was most successfully applied to graphene. Here we use that technique to find the allowed scalar and pseudomagnetic contributions to Dirac Hamiltonians in all the symmetry-enforced and symmetry-allowed orbital Dirac materials.

\section{Symmetry restrictions on Hamiltonian}
We consider the pseudomagnetic and pseudoelectric field formation in materials with Dirac nodes at high-sym\-metry points of the Brillouin zone. As shown in \cite{Manies12}, space groups of such materials and special points of their BZ are such that the little group of that point has a two-dimensional irreducible representation (which corresponds to twofold band degeneracy, i.~e., a band touching point) or in some cases a four-dimensional one and certain additional conditions are satisfied. If all irreducible representations of the little group are two- or four-dimensional, then a Dirac node is symmetry-enforced at that point. The little group $G$ of a particular point in BZ consists of the symmetry operations of the original space group which leave that point unchanged (or changed by a reciprocal lattice vector).

Electron excitations in the vicinity of the Dirac nodes are described by the Hamiltonian
\begin{equation}
H = v_x\sigma_xp_x + v_y\sigma_yp_y + v_z\sigma_zp_z.
\label{}
\end{equation}
Symmetry can impose additional restrictions on Fermi velocities $v_{x,y,z}$.

Small non-magnetic disturbance cannot open the gap in this Hamiltonian. The general form of a slightly disturbed Hamiltonian is
\begin{multline}
H = v_x\sigma_x\left(p_x-\tfrac{e}{c}A_x(\mathbf{r})\right) + v_y\sigma_y\left(p_y-\tfrac{e}{c}A_y(\mathbf{r})\right) + \\ + v_z\sigma_z\left(p_z-\tfrac{e}{c}A_z(\mathbf{r})\right) + V(\mathbf{r}).
\label{}
\end{multline}
If the disturbance is induced by small elastic strain, then the strain-induced scalar (pseudoelectric) potential $V(\mathbf{r})$ and the components of the strain-induced vector (pseudomagnetic) potential $\mathbf{A}(\mathbf{r})$ are linear combinations of components of the strain tensor, which is defined in the linear order as $u_{ij}(\mathbf{r})=\tfrac12(\partial_iu_j+\partial_ju_i)$, where $\mathbf{u}(\mathbf{r})$ is the displacement field \cite{pmfreview,GenHamGr}. Strain-induced contributions of higher order in $u_{ij}$, their derivatives, and momentum components are allowed, but they all are small if the strain is smooth and small (which is required by linear elastic strain approach) and the momentum is small, too (which is required by the Dirac treatment). For $V(\mathbf{r})$, such combination must be invariant under all elements of the little group $G$ of the point of the BZ where the Dirac node occurs (and also the elements of the space group combined with time reversal if they leave that point invariant), and $\mathbf{A}(\mathbf{r})$ must transform under elements of $G$ exactly in the same way as the momentum $\mathbf{p}$.

\section{Results for different space groups}

In Ref. \cite{Manies12}, Table I lists space groups and high-sym\-met\-ry points which always have a symmetry-enforced Dirac node, while Table II of \cite{Manies12} lists space groups and high-symmetry points which may or may not have a Dirac node (all groups in that list belong to hexagonal crystal family). Following the analogous works for graphene \cite{GenHamGr,gpmfAndo,gpmfManies} and using the database of Bilbao Crystallographic Server \cite{Bilbao,Bilbao1,Bilbao2}, we list all the groups that enforce of allow Dirac nodes at high-symmetry points in our Table~\ref{tab} together with the corresponding high-symmetry points and possible analytical forms of strain induced potentials, including scalar potential and vector potential. For scalar potentials, the contribution proportional to the trace of the strain tensor, $\sum_iu_{ii}$, is always possible, but in some cases this can be not the only possible contribution.

\begin{table*}[t]
\centering
\begin{tabular}{|c|c|c|c|}
\hline%
Space groups & Point & Scalar potential & Vector potential \\
\hline\hline%
$I4_132$ (214) & $P\left(\tfrac14,\tfrac14,\tfrac14\right)$ & $V\times(u_{xx}+u_{yy}+u_{zz})$ & $A\times(u_{yz},u_{zx},u_{xy})$ \\
\hline
$P4_132$ (213), $P4_332$ (212)      & $R\left(\tfrac12,\tfrac12,\tfrac12\right)$ & $V\times(u_{xx}+u_{yy}+u_{zz})$ & 0 \\
\hline
$I2_13$ (199) & $P$ & $V\times(u_{xx}+u_{yy}+u_{zz})$  & $A\times(u_{yz},u_{zx},u_{xy})$ \\
\hline
$P2_13$ (198) & $R$ & $V\times(u_{xx}+u_{yy}+u_{zz})$  & $A\times(u_{yz},u_{zx},u_{xy})$ \\
\hline
$I4_122$ (98) & $P$ & $V\times(u_{xx}+u_{yy})+V_zu_{zz}$  & $(Au_{yz},-Au_{zx},A_zu_{xy})$ \\
\hline
$P4_32_12$ (96), $P4_12_12$ (92) & $A\left(\tfrac12,\tfrac12,\tfrac12\right)$ & $V\times(u_{xx}+u_{yy})+V_zu_{zz}$  & $A\times(u_{yz},-u_{zx},0)$ \\
\hline
$I2_12_12_1$ (24) & $W\left(\tfrac34,\tfrac{\bar{1}}{4},\tfrac{\bar{1}}{4}\right)$ & $V_xu_{xx}+V_yu_{yy}+V_zu_{zz}$  & $(A_xu_{yz},A_yu_{zx},A_zu_{xy})$ \\
\hline
$P2_12_12_1$ (19) & $R$ & $V_xu_{xx}+V_yu_{yy}+V_zu_{zz}$  & $(A_xu_{yz},A_yu_{zx},A_zu_{xy})$ \\
\hline\hline
$P622,P6_{1,5,2,4,3}22$ (177--182) & $K\left(\tfrac13,\tfrac13,0\right)$& $V\times(u_{xx}+u_{yy})+V_zu_{zz}$  & $A\times(u_{xx}-u_{yy},-2u_{xy},0)$\\
\hline
$P622,P6_{2,4}22$ (177,180,181) & $H\left(\tfrac13,\tfrac13,\tfrac12\right)$& $V\times(u_{xx}+u_{yy})+V_zu_{zz}$  & $A\times(u_{xx}-u_{yy},-2u_{xy},0)$\\
\hline
$P321,P3_{1,2}21$ (150,152,154)  &$K,H$&$V\times(u_{xx}+u_{yy})+V_zu_{zz}$&$A\!\times\!(u_{xx}\!-\!u_{yy},-2u_{xy},0)+A_z\!\times\!(u_{yz},-u_{zx},0)$ \\
\hline
\end{tabular}
\caption{\label{tab}
Strain-induced potentials for symmetry-enforced (top part) and symmetry-allowed (bottom part) Dirac nodes at high-symmetry points of BZ for given space groups.
}
\end{table*}

For example, consider the space group $I4_132$ (No.~214 in \cite{itcbook}). It must have a Dirac node at $P$ point of its Brillouin zone. This space group is generated by following symmetry operations:
\begin{eqnarray}
\left( \begin{array}{rrr|r} -1&0&0&1/2\\0&-1&0&0\\0&0&1&1/2 \end{array} \right), \left( \begin{array}{rrr|r} -1&0&0&0\\0&1&0&1/2\\0&0&-1&1/2 \end{array} \right), \nonumber\\ \left( \begin{array}{rrr|r} 0&0&1&0\\1&0&0&0\\0&1&0&0 \end{array} \right), \left( \begin{array}{rrr|r} 0&1&0&3/4\\1&0&0&1/4\\0&0&-1&1/4 \end{array} \right).
\end{eqnarray}
The symmetry operation consists of an orthogonal transformation represented by a matrix and a translation represented by a vector. Only the matrices are relevant for our consideration.

All but the last of these symmetry operations also enter the little group of $P$. It can be seen that the symmetry enforces $v_x=v_y=v_z$. The only possible combination of the strain tensor components which is invariant under little group operations is the trace of the strain tensor, and the only possible vector made of strain tensor components which transforms in the same way as the momentum vector is $(u_{yz},u_{zx},u_{xy})$.

\begin{figure}[t]
\centering
\resizebox{0.8\columnwidth}{!}{\includegraphics{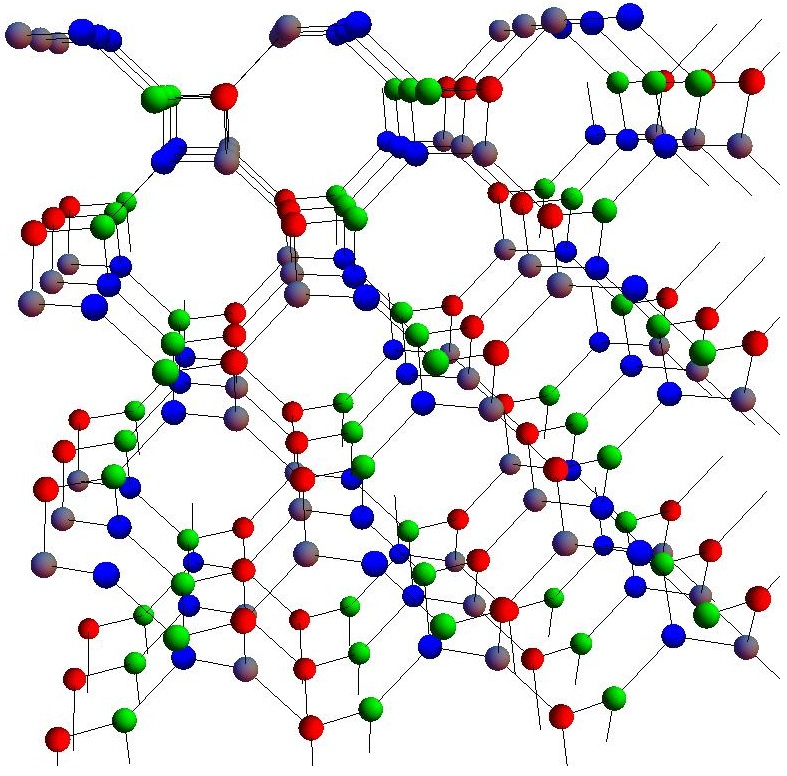}}
\caption{\label{fig:k4}
The $(10,3)$-$a$ crystal, one of the simplest crystal structures having space group $I4_132$. The crystal has 4 primitive bcc sublattices, shown in different colors.
}
\end{figure}

This can be confirmed by an explicit example. One of the simplest structures having space group $I4_132$ is the $(10,3)$-$a$ crystal (Fig.~\ref{fig:k4}), also known as triamond, or the $K_4$ crystal (see \cite{K4} and references therein). Two Dirac nodes with different energies can be observed in the band structure given by a four-band $s$-orbital tight-binding model of that crystal, considered in \cite{Manies12} and \cite{Tsuchiizu}; that system is not a Dirac semimetal though, because Dirac fermions coexist with other electron excitations, which are much more numerous. The corresponding Fermi velocity is $v_\mathrm{F} = \tfrac{1}{2\sqrt3}at/\hbar$, where $a$ is the lattice constant and $t$ is the hopping integral. If we assume that hoppings are modified by strain with $\beta=-\frac{a}{t}\frac{\partial t}{\partial a}=\text{const}$, then the strained Dirac Hamiltonian acquires the scalar potential $V(\mathbf{r}) = \frac{1}{\sqrt3}\beta t\times(u_{xx}+u_{yy}+u_{zz})$ and the vector potential $e\mathbf{A} = \frac{c}{v_\mathrm{F}}\beta t\times(u_{yz},u_{zx},u_{xy})$, which fully fits the form predicted by symmetry.

Table~\ref{tab} shows a diversity of expressions for potentials. For the group $P4_132$ and its enantiomorphic pair $P4_332$, the PMF is forbidden. Not surprisingly, for some of the hexagonal groups, the PMF potential is the same as for the case of graphene. The factors in the expansions given in the table cannot be predicted from symmetry.

In all the cases listed, apart from a Dirac node in a high-symmetry point, there is another Dirac node at the opposite corner of the Brillouin zone. When the high-symmetry point is time-reversal-invariant (it happens for simple cubic lattices, groups from the top part of Table~\ref{tab} having $P$ in their Hermann--Maugin symbol), there are two kinds of Dirac fermions living at the same Dirac node (or, two Dirac nodes at the same point of the Brillouin zone and the same energies) \cite{Manies12}. Thus, the second Dirac node and the corresponding valley is always related to the first one by time reversal. It changes the signs of momenta but does not affect strain tensor. Consequently, the strain-induced scalar potential has the same sign for two valleys related by time reversal, while the PMF potential has different sign. This is consistent with the fact that strain cannot violate time-reversal symmetry.

This, in turn, means that PMF can be used to control valley currents, which can be applied to 3D valleytronics \cite{ChinValley}. In Weyl and multi-Weyl semimetals, the topological responses from strain-induced axial fields manifest themselves in various physical effects (including chiral anomaly \cite{PRX1}), allowing to control chirality accumulation, which opens further opportunities for valleytronics \cite{TopResp}. PMF in Dirac materials can also manifest itself in pseudo-Landau levels \cite{PRX2}.

Three-dimensional Dirac (and Weyl) semimetals usually have Dirac (Weyl) nodes not at high-symmetry points, but rather at high-symmetry lines or planes of their Brillouin zones, and/or spin-orbit coupling is present in these materials \cite{DiracReview14,DSMAnnuRev,DSMJPCM,NewRev}. However, Dirac nodes described in \cite{Manies12} can be found in the band structure of real materials. In particular, the cubic gauche nitrogen (cg-N), a high-energy-density nitrogen allotrope \cite{cgNexp,cgNexpNew}, has space group $I2_13$, and its band structure indeed shows Dirac nodes at the $P$ point of the BZ, including one just above the Fermi level \cite{cgNbands}. Although cg-N is not a Dirac semimetal, the density of states at the Dirac node energy is little enough to assume that Dirac fermion dynamics can be seen in \mbox{cg-N}. One possible way to see the effect of the PMF is to make a commensurate two-dimensional thin slice including the direction of the $P$ point, e.~g. parallel to the $(1\bar{1}0)$ plane, introduce orthogonal coordinates $(q,z)$ in it, apply an appropriate strain profile generating a uniform PMF in accordance with Table~\ref{tab}, e.~g. in-plane deformation $\mathbf{u}=(z^2/L,0)$ with $L$ being a parameter of dimension of length, move the Fermi level to the Dirac node, and observe the pseudo-Landau levels.

\section{Conclusion}
The effect of strain upon materials with Dirac nodes at high-symmetry points of Brillouin zone have been studied with symmetry considerations. The results shown in Table~\ref{tab} indicate that the form of the dependence of the strain-induced scalar and vector potentials differs for different space groups; scalar potential is always present, while the vector pseudomagnetic field potential is present in all cases except for space groups $P4_132$ and $P4_332$. As the pseudomagnetic field has different sign for two valleys, it can be used for 3D valleytronics. The pseudomagnetic field can manifest itself in cubic gauche nitrogen.

\section*{Acknowledgements}
The work was supported by RFBR [grant number 17-02-01134]. A.\,D.\,Z. was supported by the ``PhD Student'' grant of Foundation for the advancement of theoretical physics and mathematics ``Basis''. Yu.\,E.\,L. was supported by Program of Basic Research of HSE.

\bibliography{jmmm_paper}

\end{document}